\documentclass[twocolumn,showpacs,preprintnumbers,amsmath,amssymb,prl]{revtex4}
\usepackage{amsbsy}
\usepackage{graphicx,color}
\usepackage{amsfonts}
\usepackage{amsmath}
\usepackage{amssymb}
\date{}                                           % Activate to display a given date or no date

\newcommand{\ds}{ _{\downarrow}}
\newcommand{\us}{ _{\uparrow}}
\newcommand{\up}{\uparrow}
\newcommand{\down}{\downarrow}

\begin{document}
\draft
\title{Atom-dimer p-wave resonance for fermionic mixtures with different masses}
\author{R. Combescot$^{(a),(b)}$ and  X. Leyronas$^{(a)}$}
\address{(a) Laboratoire de Physique Statistique, Ecole Normale Sup\'erieure, UPMC  
Paris 06, Universit\'e Paris Diderot, CNRS, 24 rue Lhomond, 75005 Paris,  
France.}
\address{(b) Institut Universitaire de France,
103 boulevard Saint-Michel, 75005 Paris, France.}
\date{Received \today}
\pacs{03.75.Ss, 03.65.Nk, 34.50.Cx, 67.85.Lm}

\begin{abstract}
We show that, near a Feshbach resonance, a strong p-wave resonance is present at low energy in atom-dimer scattering 
for  $^6$Li-$^{40}$K fermionic mixtures. This resonance is due to a virtual bound state, in the atom-dimer system, which is 
present at this low energy. When the mass ratio between the two fermionic elements is increased, this virtual bound state 
goes to a known real bound state which appears when the mass ratio reaches 8.17. This resonance should affect a number 
of physical properties. These include the equation of state of unbalanced mixtures at very low temperature but also the
equation of state of balanced mixtures at moderate or high temperature. The frequency and the damping of collective modes
should also provide a convenient way to evidence this resonance. Finally it should be possible to modify the effective mass
of one the fermionic species by making use of an optical lattice. This would allow to study the strong dependence of the
resonance as a function of the mass ratio of the two fermionic elements.
\end{abstract}
\maketitle

One of the most remarkable features of the physics of ultracold gases is the simplicity of the effective interaction
which is completely characterized by the scattering length between the various involved atoms.
For example in fermionic gases involving only atoms in the two lowest hyperfine states of a single element,
such as $^6$Li or $^{40}$K, the interaction is fully described by the scattering length $a$ between atoms belonging to these two
different hyperfine states \cite{gps}. Remarkably this simplicity extends even to apparently more complex situations.
Indeed on the BEC side of a Feshbach resonance, where $a>0$, fermionic atoms in two different hyperfine
states may form molecules, or dimers, with binding energy $E_b$. One might expect that the formation of
these bosons could lead to more complex physics, expected for a molecular gas. Nevertheless the $T=0$ equation of state on this BEC
side is very well described \cite{gps} in a wide domain of density by the Lee-Huang-Yang \cite{lhy} equation of state
for these composite bosons \cite{lc}, making merely use of the dimer-dimer scattering length $a_4=0.6 \, a$
between these bosons \cite{pss,bkkcl}. This is as if these bosons were elementary.

In this paper we show that this simplicity is partially linked to the fact that the two atomic species have equal masses. 
It should be lost in the case of mixtures of atomic fermions corresponding to different elements, such as mixtures of 
$^6$Li and $^{40}$K toward which the interest has recently turned. Indeed new interesting physics is expected
for large mass ratio \cite{pao,parish,blume,baranov,ops,brs, diener}. Quite recent theoretical work \cite{ggsc,bdgc}
has also focused on the case where the particle number is small and multiparticle bound states may occur. Experimentally 
much progress has been made in exploring these mixtures \cite{munich,innsb,ens} and degeneracy has already been reached. 
In particular a Feshbach resonance, which is fairly broad and reasonably stable, has been identified near $155$G \cite{inns}.
In the following we will assume that we are near such a Feshbach resonance and ignore any stability problem.
We will follow the standard use of refering to the two different atomic species as $\up$ and $\down$ atoms, even if there is
no relation between this notation and the physical spin of these atoms. We note $a$ the scattering length between these atoms
and assume that a zero-range interaction is a good enough approximation so that the universal regime applies. Finally we ignore any trap effect.

A first sign of more complex physics is the increase \cite{pss1} (for a given $a$) of the dimer-dimer scattering length with the mass ratio
$m\up/m\down$. However the most important point by far is the appearance of new bound states when this mass ratio increases.
These bound states appear already at the level of the three-body problem and for the fermionic situation they have been
studied in great details by Kartavtsev and Malykh \cite{karma}, for two identical fermions with mass $m\up$ and a different fermion
with mass $m\down$. They are not present in the s-wave channel and appear only
for angular momenta $\ell \ge 1$. Among them are the well-known Efimov \cite{efimov} states, which are quite remarkable for
their spectrum structure with binding energy going up to infinity when the range of the interaction goes to zero.
For $\ell =1$ the Efimov states appear only \cite{efi1,karma} for a mass ratio $m\up / m\down = 13.6$. However there are already bound states
for smaller mass ratios, which are not Efimov states. A first bound state \cite{karma} appears for mass ratio $m\up / m\down = 8.17$.

Actually we show in this paper that very important modifications in the fermion-dimer scattering properties arise for quite lower mass ratios.
The basic reason is that, even before the mass ratio threshold for bound states is reached, virtual bound states are already present.
For nonzero angular momenta, in particular $\ell=1$, the centrifugal barrier acts to inhibit their decay and provide them with
fairly long lifetime. Hence they are not much different from real bound states. Moreover, since their energy is positive, they
give rise to resonances, which are strong when they are located at low energy. We find that these effects are already important
for the mass ratio $m\up / m\down = 6.64$ corresponding to the mixtures of $^6$Li and $^{40}$K, which is probably the most
investigated experimentally. We note that these virtual bound states could play a very important role in the three-body decay
and in the stability of these mixtures, since they correspond to physical situations where three atoms stay close together for
a fairly long time, increasing in this way the probability of three-body decay processes. We observe that the effect of the mass ratio
can be studied experimentally, since the use of optical lattices allows a manipulation of the effective value of the atomic mass.
These lattices could also be used to study and possibly remove instability problems.

\begin{figure}
\centering
%\hspace*{20mm}
{\includegraphics[width=\linewidth]{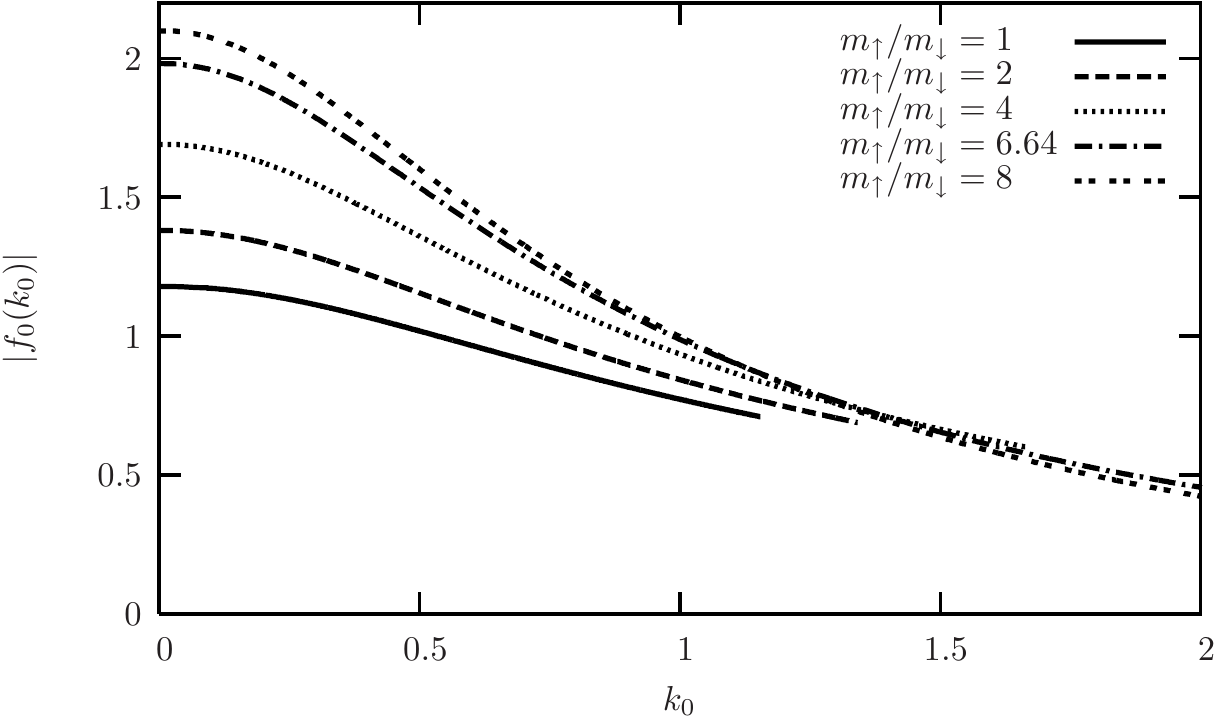}}
\caption{Modulus of the $\ell=0$ scattering amplitude $|f_0(k_0)|$ as a function of $k_0$, for mass ratios
$m\us / m\ds=1, 2, 4, 6.64$ and $8$. The ratio $m\us / m\ds=6.64$ corresponds to the $^6$Li - $^{40}$K
mixture. All these quantities are in reduced units, equivalent to set $a=1$.}
\label{fig1}
\end{figure}

In this paper we are interested in mass ratios below the value $8.17$, corresponding to the appearance of the first p-wave bound state.
We consider only the s-wave and p-wave components of the full scattering amplitude,
since the higher angular momentum components are expected to give very small contributions
to the total scattering.
We show that the p-wave contribution becomes very strong, due to quasi-resonance, when the mass ratio approaches $8.17$.

This analysis is done quite conveniently by making
use of the integral equation formulation of this dimer-fermion scattering problem first worked out by 
Skorniakov and Ter-Martirosian\cite{stm}. We use the full Skorniakov and Ter-Martirosian equations, 
corresponding to different fermion masses, for nonzero kinetic energy in the center of mass referential.

We introduce the total mass of the dimer $M=m\up +m\down$, its reduced mass 
$\mu =m\us m\ds /(m\us + m\ds)$ and its binding energy $E_b=1/(2\mu a^2)$.
We work in the center of mass referential for which the total momentum $\textbf{P}={\bf 0}$.
The corresponding total energy is $E=-1/(2\mu a^2)+k_0^2/2\mu _T$
where $\mu _T$ is the atom-dimer reduced mass $\mu _T = m\us M /(m\us + M)=m\us (m\us + m\ds)/(2 m\us +m\ds)$,
and ${\bf k}_0$ the outgoing momentum of the $\up$ fermion.
The masses enter only through the combinations $R=\mu /\mu _T$, $R'=2 \mu /m\ds$.

\begin{figure}
\centering
%\hspace*{20mm}
{\includegraphics[width=\linewidth]{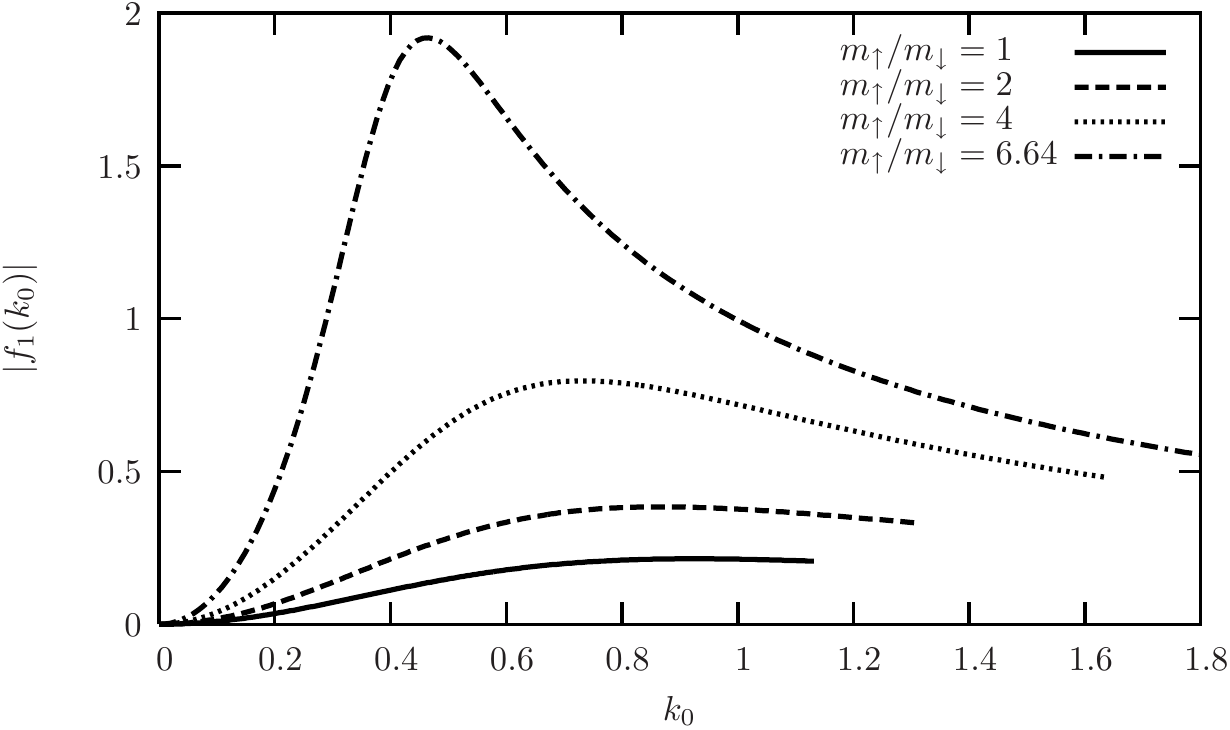}}
\caption{Modulus of the $\ell=1$ scattering amplitude $|f_1(k_0)|$ as a function of $k_0$, for mass ratios
$m\us / m\ds=1, 2, 4$ and $6.64$. The ratio $m\us / m\ds=6.64$ corresponds to the $^6$Li - $^{40}$K mixture.}
\label{fig2}
\end{figure}

The unknown function $a_{3}({\bf k},{\bf k}_0)$ in the Skorniakov-Ter-Martirosian integral equation is, within a multiplicative constant,
the $T$-matrix for this scattering problem. Its s-wave component $a_{3 \rm s}(k,k_0)$ is given by $a_{3 \rm s}(k,k_0)
=(1/4\pi ) \int d\Omega_{\bf k} a_{3}({\bf k},{\bf k}_0)$, while its p-wave component reads
$a_{3 \rm p}(k,k_0)=(1/4\pi ) \int d\Omega_{\bf k} P_1({\bf k},{\bf k}_0) a_{3}({\bf k},{\bf k}_0)$ where
$P_1({\hat {\bf k}},{\hat {\bf k}}_0)={{\bf k}} \cdot {\bf k}_0/kk_0$ is the $\ell=1$ Legendre polynomial.
The corresponding s-wave and p-wave scattering amplitudes are given by $a_{3 \rm s}(k_0,k_0)$
and $a_{3 \rm p}(k_0,k_0)$ respectively. The integral equations satisfied by $a_{3 \rm s}(k,k_0)$
and $a_{3 \rm p}(k,k_0)$ are obtained by projecting the general Skorniakov and Ter-Martirosian equation
on the $\ell=0$ and $\ell=1$ components. The resulting equations are quite similar.
In order to make the equations simpler, we use systematically reduced variables:
all the wavevectors are actually reduced (dimensionless) wavevectors, expressed in units of $1/a$,
and similarly $a_{3 \rm s,p}(k,k_0)$ are reduced scattering amplitudes expressed in units of $a$.

Introducing the notations:
\begin{eqnarray}\label{}
\alpha (k,q)&=&1-Rk_0^2+k^2+q^2 \\ \nonumber
\beta (k,q)&=&R'kq
\end{eqnarray}
we define the kernels:
\begin{eqnarray}\label{}
K_{\rm s}(k,q)=\frac{1}{2\beta(k,q)}\ln \left|\frac{\alpha (k,q)+\beta (k,q)}{\alpha (k,q)-\beta (k,q)}\right|
\end{eqnarray}
and
\begin{eqnarray}\label{}
K_{\rm p}(k,q)=\frac{1}{\beta (k,q)}-\frac{\alpha (k,q)}{2\beta^2 (k,q)}\ln \left|\frac{\alpha (k,q)+\beta (k,q)}{\alpha (k,q)-\beta (k,q)}\right|
\end{eqnarray}
Then the equations for the s-wave and p-wave components are:
\begin{eqnarray}\label{inteq}
\!\!R\;\frac{a_{3 \rm s,p}(k,k_0)}{1+\sqrt{1+ R\,(k^2-k_0^2)}}\!\!&=&\!\!K_{\rm s,p}(k,k_0) \nonumber \\
- \frac{2}{\pi} \int_{0}^{\infty} dq\!\!&&\!\!\frac{q^2 a_{3 \rm s,p}(q,k_0)}{q^2-k_0^2-i\delta}\;K_{\rm s,p}(k,q)
\end{eqnarray}

We consider first the results for the s-wave component. Since we are not interested in the phase shift,
we restrict ourselves to the modulus $|f_0(k_0)|=|a_{3 \rm s}(k_0,k_0)|$ of the scattering amplitude.
Here as well as in the following we investigate only negative values of the energy $E$ in order
to avoid consideration of inelastic processes, leading to the dimer breaking. This implies $k_0^2 <1/R$.
The results for $|f_0(k_0)|$ in terms of $k_0$ are given in Fig.\ref{fig1}, for the mass ratios $m\us / m\ds=
1, 2, 4, 8$. We have also inserted the result for $m\us / m\ds=6.64$ corresponding to the $^6$Li - $^{40}$K
mixture.

The $k_0=0$ value is equal to the scattering length, and increases from $1.179$ corresponding to $m\us = m\ds$ to
$2.099$ for $m\us = 8 m\ds$. This known fairly slow increase of the atom-dimer scattering length with the mass ratio
is the dominant feature of Fig.\ref{fig1}, since otherwise all the results are of order unity for $k_0=1$ and keep decreasing for
increasing $k_0$.

\begin{figure}
\centering
%\hspace*{20mm}
{\includegraphics[width=\linewidth]{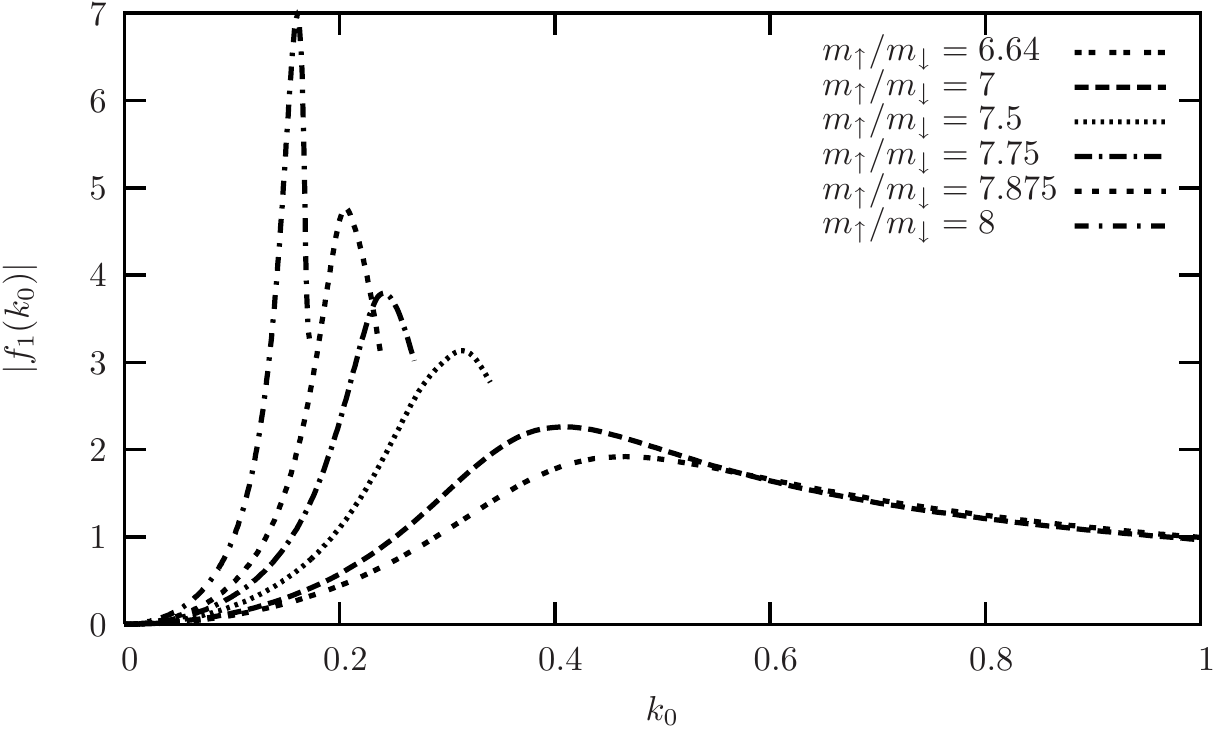}}
\caption{Modulus of the $\ell=1$ scattering amplitude $|f_1(k_0)|$ as a function of $k_0$, for mass ratios
$m\us / m\ds=6.64$ (corresponding to the $^6$Li - $^{40}$K mixture), $7, 7.5, 7.75, 7.875$ and $8$.}
\label{fig3}
\end{figure}

We turn now to the p-wave component results. We first give in Fig.\ref{fig2} the results for mass ratios
$m\us / m\ds=1, 2, 4$ and $6.64$, stopping at the value corresponding to the $^6$Li - $^{40}$K
mixture. For equal masses the p-wave component is always small, as it could be
expected, and it can clearly be omitted when
one deals with the atom-dimer scattering properties. Hence the atom-dimer vertex can, to a large extent,
be taken as a constant proportional to the atom-dimer scattering length. No complication is thus expected
to arise from this side when many-body properties will be investigated.  The situation is qualitatively similar for $m\us / m\ds= 2$ 
and even $4$: the p-wave component is fairly small and can be neglected compared to the s-wave contribution. 
However, we see that this p-wave
component rises steeply when the mass ratio goes from $4$ to $6.64$. This last ratio is at the border of
the resonant domain. 

\begin{figure}
\centering
%\hspace*{20mm}
{\includegraphics[width=\linewidth]{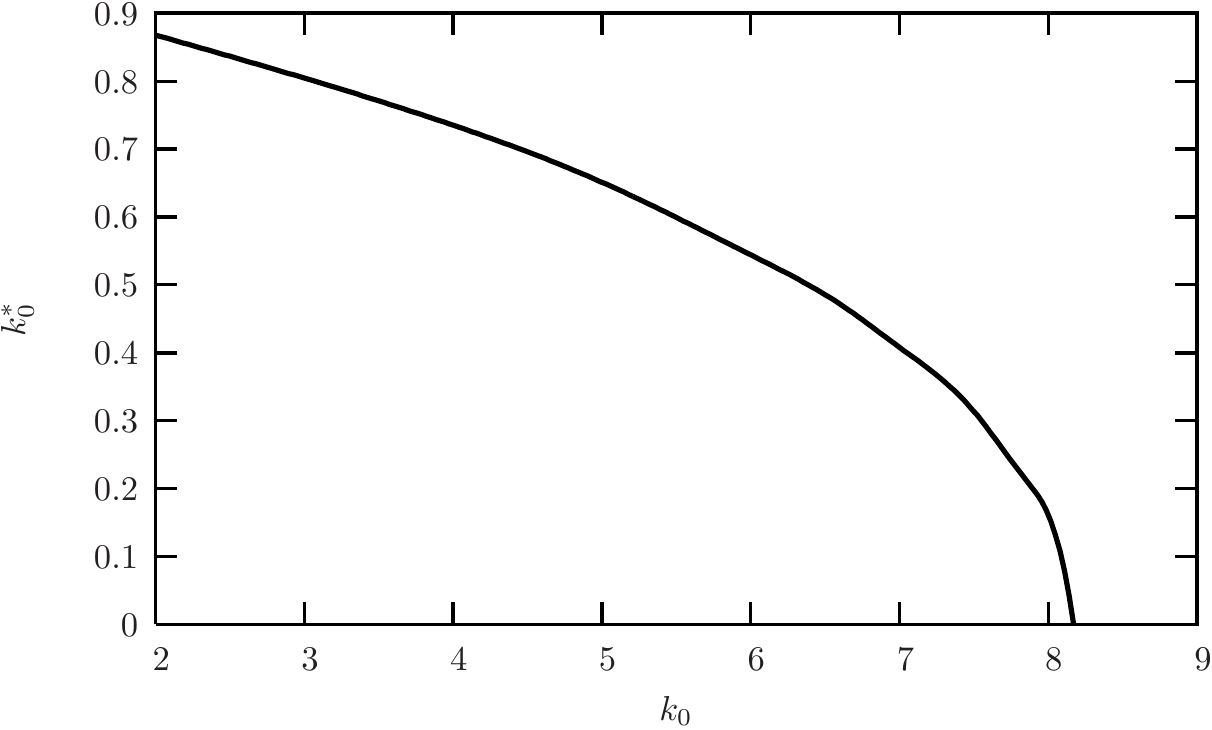}}
\caption{Location of the maximum $k_0^*$ of $|f_1(k_0)|$ as a function of mass ratio $m\us / m\ds$.}
\label{fig4}
\end{figure}

This domain is now displayed in Fig.~\ref{fig3},  where the results for mass ratios
$m\us / m\ds=6.64, 7., 7.5, 7.75, 7.875$ and $8.$ are plotted.
We see that the hump, present for $m\us / m\ds=6.64$ around $k_0 \simeq 0.45$, develops into a strong resonance
at lower and lower energy when $m\us / m\ds$ is increased. Although we have sampled mass ratios which are very
close to each other, we see that the resonance grows very rapidly with increasing mass ratio. The physical origin
is clearly the development of a virtual bound state at positive energy. The lifetime of this state is directly related to the width
of the resonance. Since the resonance peak gets quickly narrower, the lifetime grows very rapidly with increasing mass.
The link with the bound state which appears \cite{karma} at zero energy for $m\us / m\ds=8.17$ is confirmed if we
look at the position of the resonance peak as a function of the mass ratio. This is shown in Fig.\ref{fig4}.
We see that the position $k_0^*$ of the resonance peak extrapolates to zero for a mass ratio which is in close vicinity of $8.17$.
More precisely when the resonance peak reaches $k_0^*=0$, it will be infinitely sharp, corresponding to an infinite value
for $|f_1(k_0=0)|$. Hence in this case $a_{3 \rm p}(k,0)$ will be infinitely large, which implies that the homogeneous
Eq.(\ref{inteq}) (i.e. without the term $K_{\rm p}(k,0)$ in the right-hand side) has a solution for zero energy $k_0=0$.
This is just stating the well-known result that bound states are solutions of the homogeneous integral equation
corresponding to Eq.(\ref{inteq}). It is easy to find the lowest $m\us / m\ds$ for which this homogeneous integral equation
has a solution. We find that this occurs for $m\us / m\ds=8.172$ in full agreement with Kartavtsev and Malykh \cite{karma}.

\begin{figure}
\centering
%\hspace*{20mm}
{\includegraphics[width=\linewidth]{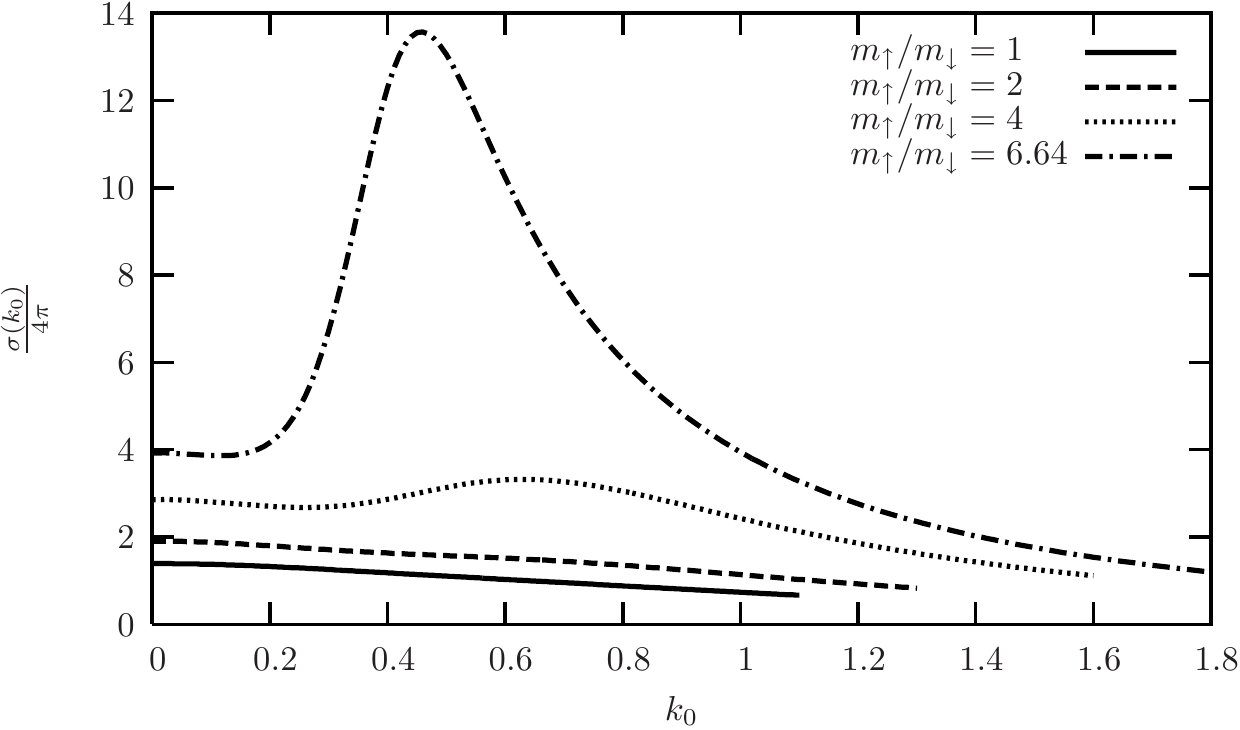}}
\caption{Total scattering cross section (divided by $4\pi $) Eq.(\ref{sigm}) as a function of $k_0$, for mass ratios
$m\us / m\ds=1, 2, 4$ and $6.64$.}
\label{fig5}
\end{figure}

\begin{figure}
\centering
%\hspace*{20mm}
{\includegraphics[width=\linewidth]{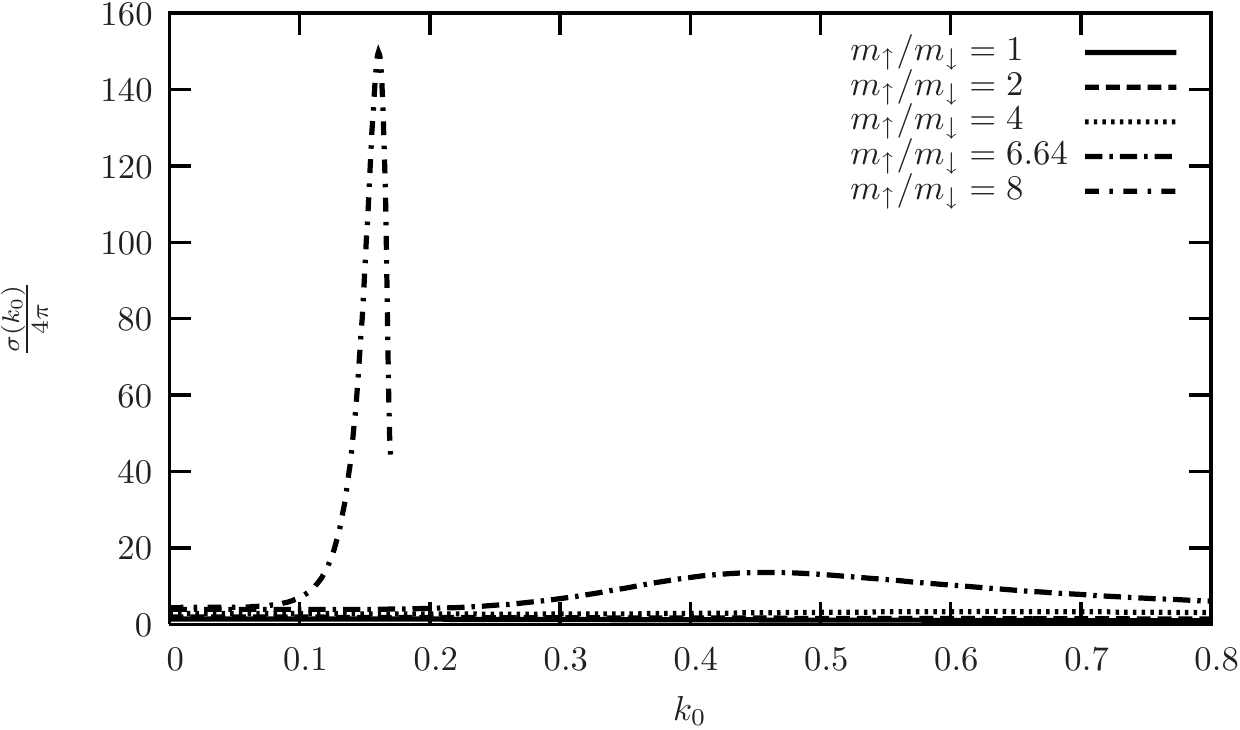}}
\caption{Total scattering cross section (divided by $4\pi $) Eq.(\ref{sigm}) as a function of $k_0$, for mass ratios
$m\us / m\ds=1, 2, 4$ and $6.64$ and $8$.}
\label{fig5bis}
\end{figure}

Let us now come to the total scattering cross section $\sigma(k_0)$, which is likely to be the easiest direct physical quantity to measure
experimentally. Since we neglect angular momenta higher than $\ell=1$, we have for this cross section:
\begin{eqnarray}\label{sigm}
\frac{\sigma(k_0)}{4\pi }=\!\sum_{\ell} (2 \ell +1)|f_{\ell}(k_0)|^2\!=\!|f_0(k_0)|^2+3|f_1(k_0)|^2
\end{eqnarray}
The result is plotted in Fig.\ref{fig5} for $m\us / m\ds=1, 2, 4$ and $6.64$. We see that, while up to $m\us / m\ds= 4$,
the cross section is almost featureless, it display a strong resonance for $m\us / m\ds=6.64$ which corresponds to the
experimental value for $^6$Li - $^{40}$K mixture. Naturally when the mass ratio is further increased, this resonance
becomes even stronger. This is shown in Fig.~\ref{fig5bis} where we display also the case of $m\us / m\ds=8$. The
resulting resonance dwarfs the preceding one.

This strong resonance in the atom-dimer scattering cross section for $^6$Li - $^{40}$K mixture should imply a number 
of experimentally observable consequences. The more natural way
to evidence it would be to measure the collision properties between a $^{40}$K cloud and a $^{40}$K -$^6$Li dimer
cloud, in a way analogous to the recent experiments on $^6$Li \cite{zwierlein}. However this might not be so easy to perform.
Measuring the equation of state of a mixture is on the other hand a fundamental question which should be easier to answer.
The resonance should affect in an important way the low temperature equation of state of a mixture of $^{40}$K cloud and a $^{40}$K 
-$^6$Li dimer, arising from an unbalanced mixture of $^{40}$K and $^6$Li atoms on the BEC side of the Feshbach
resonance. This will not appear at the mean field level since the angular average will cancel the effect because this
is a p-wave resonance. But going to next order in $^{40}$K, analogously to second order perturbation theory, there will be no
cancellation. Hence one should see a strong dependence on the $^{40}$K density. One expects also a strong density
dependence to arise when the Fermi wavevector of the $^{40}$K Fermi sea reaches the wavevector corresponding
to the resonance, since for lower wavevectors the scattering is essentially negligible. One should see something
similar to a threshold effect in density. We note that the resonance should also have an important effect even on a
balanced mixture of $^{40}$K and $^6$Li, on the BEC side, when the temperature is raised. Indeed the dimers will
be partially broken by thermal excitation which provides a natural source of free $^{40}$K atoms, and effects analogous
to the ones arising in the unbalanced mixture. Similarly, going to high temperature, we expect this resonance to have
a marked effect on the virial coefficients since these are systematically related to the three-body problem \cite{xav}.
Naturally the resonance is also expected to affect strongly
the transport properties. Both the effects on the equation of state and on the transport properties should appear
in the frequency and the damping of the collective modes \cite{gps}.

Another way to understand the effect of this resonance on the $^6$Li - $^{40}$K mixtures is to remark that
the long lived virtual bound states responsible for it should behave in many respects in a way analogous to real bound states.
Hence on the BEC side the physical description should involve not only free fermions and dimers, but
also the existence of trimers. Obviously all the physical properties should be affected by the presence
of this additional fermion species. A related point is that the existence of these trimers should affect the
dimer-dimer scattering properties. In other words we have only explored the simpler fermion-dimer
scattering properties, but we expect that the dimer-dimer scattering amplitude will display related
energy structure, and they should not be so complicated to explore with the methods of our preceding
work \cite{bkkcl}.

Finally it is worth stressing that the use of optical lattices should provide a very convenient and powerful way
to explore the resonance we have pointed out. Indeed the natural mass ratio of $6.64$ between $^{40}$K 
and $^6$Li atoms happens to be just at the border of the strongly resonating mass ratio domain. A rather weak
optical lattice could be used to slightly tune the effective mass of $^{40}$K or $^6$Li (by making an appropriate
choice of the light frequency, only one atomic species is affected), producing strong modification of the resonance 
and hence of the physical properties of the mixture. In this way one could go into the strongly resonating domain, 
or even reach the threshold for the appearance of real bound states. Or one could go in the other direction and basically
get rid of the resonance, which would allow to prove that it is responsible for specific physical properties of
the mixture. More specifically the 3D potential $V_{\rm opt}({\bf r})=s E_R \left[\sin^2(K_0x)+\sin^2(K_0y)+\sin^2(K_0z)\right]$
with $E_R=K_0^2/2m$ provides, within second order perturbation theory, an effective $m^*$ given by $m/m^*=1-s^2/32$.
Increasing the $^{40}$K effective mass to reach the bound state ratio $8,17$ requires $s=2.45$ which corresponds
to a fairly weak optical potential. This simple picture of effective mass modification works only if the size of the
involved objects (dimer, trimer) is large compared to the optical wavelength $\lambda$. Taking $\lambda \sim 500\, {\rm nm}$
this should be a valid approximation in the vicinity where the scattering length becomes quite large. Even if corrections
to this simple picture are necessary, the qualitative physical trends should remain valid.

The authors wish to thank F. Alzetto for useful discussions.

\end{document}